\documentclass[journal]{IEEEtran}

\usepackage{cite}
\usepackage{amsmath,amssymb,amsfonts}
\usepackage{algorithmic}
\usepackage{graphicx}
\usepackage{textcomp}
\usepackage{amssymb}
\usepackage{pifont}
\usepackage{array}
\usepackage[T1]{fontenc}

\usepackage{hyperref}
\usepackage{pifont}
\newcommand{\xmark}{\ding{55}}
\newcommand{\cmark}{\ding{51}}

\usepackage{enumitem}
\usepackage{multirow}

\ifCLASSINFOpdf
\else
\fi

\begin{document}

\title{Mammo-Clustering: Context Clustering based Multi-view Tri Level Information Fusion for Lesion Location and Classification in Mammography}

\author{
\thanks{This work is partially supported by grants from the Clinical Research Project of the First Affiliated Hospital of Shenzhen University (2023YJLCYJ019)}
Shilong Yang, Chulong Zhang \IEEEmembership{Member, IEEE}, Xiaokun Liang, \IEEEmembership{Member, IEEE},
\thanks{Shilong Yang, Chulong Zhang, Xiaokun Liang and Yaoqin Xie are with the Shenzhen Institutes of Advanced Technology, Chinese Academy of Sciences, 1068 Xueyuan Avenue, Shenzhen University Town, China, 518055.}
Qi Zang, \thanks{Qi Zang is with the Qingdao University, Qingdao, China, 266000.}
Juan Yu, Liang Zeng, Xiao Luo, Yexuan Xing, Xin Pan, Qi Li,
\thanks{Juan Yu, Liang Zeng, Xiao Luo, Yexuan Xing, Xin Pan, Qi Li are with the Department of Radiology, The First Affiliated Hospital of Shenzhen University, Health Science Center, Shenzhen Second People's Hospital, 3002 SunGangXi Road, Shenzhen, 518035, China.}
Linlin Shen, \IEEEmembership{Senior, IEEE},
\thanks{Linlin Shen is with the Shenzhen University, Shenzhen, China, 518060.}
and Yaoqin Xie\thanks{These authors contributions are equal: Shilong Yang and Chulong Zhang}
\thanks{corresponding authors: Linlin Shen and Yaoqin Xie (e-mail: llshen@szu.edu.cn and yq.xie@siat.ac.cn)}
}

\maketitle

\begin{abstract}
Breast cancer is a significant global health issue, and the diagnosis of breast cancer through imaging remains challenging. Mammography images are characterized by extremely high resolution, while lesions often occupy only a small portion of the image. Down-sampling in neural networks can easily lead to the loss of microcalcifications or subtle structures. To tackle these challenges, we propose a Context Clustering based triple information fusion framework. First, in comparison to CNNs or transformers, we observe that Context clustering methods are (1) more computationally efficient and (2) better at associating structural or pathological features. This makes them particularly well-suited for mammography in clinical settings. Next, we propose a triple information fusion mechanism that integrates global, feature-based local, and patch-based local information.
The proposed approach is rigorously evaluated on two public datasets, Vindr-Mammo and CBIS-DDSM, using five independent data splits to ensure statistical robustness. Our method achieves an AUC of \(0.828 \pm 0.020\) on Vindr-Mammo and \(0.805 \pm 0.020\) on CBIS-DDSM, outperforming the second best method by 3.5\% and 2.5\%, respectively. These improvements are statistically significant (\(p<0.05\)), highlighting the advantages of the Context Clustering Network with triple information fusion. 
Overall, our Context Clustering framework demonstrates strong potential as a scalable and cost-effective solution for large-scale mammography screening, enabling more efficient and accurate breast cancer detection. Access to our method is available at \href{https://github.com/Sohyu1/Mammo-Clustering}{https://github.com/Sohyu1/Mammo-Clustering}.
\end{abstract}


\section{Introduction}
\label{sec:introduction}
\IEEEPARstart{B}{reast} 
cancer, as the most prevalent malignancy among women, has surpassed cardiovascular diseases as the leading cause of premature death in women worldwide \cite{4} \cite{5}. Nevertheless, breast cancer is particularly amenable to effective prevention and treatment strategies \cite{18}. Early detection is critical for reducing mortality rates and improving patient prognosis \cite{6} \cite{19}. It facilitates the implementation of less invasive and more targeted treatment options, thereby alleviating the physical and psychological burden on patients.

Furthermore, studies have shown that early breast cancer screening using mammography can significantly reduce mortality rates by up to 20\% \cite{1}.
Mammography is a low-dose, non-invasive X-ray imaging technique \cite{29} that plays a crucial role in the early detection of breast cancer by identifying tumors too small to be palpated, thereby facilitating timely intervention. 

One aspect of the specificity of the mammography issue is that, as a multi-view imaging technique, mammograms are typically acquired from the craniocaudal (CC) and mediolateral oblique (MLO) angles of both the left and right breasts. From a given perspective, the symmetry between the left and right breasts also serves as a critical diagnostic criterion in clinical practice. Consequently, employing a multi-view learning strategy can leverage the complementary information provided by different imaging angles, thereby enhancing classification performance \cite{36} \cite{65}.

Multi-view learning is a machine learning paradigm that leverages multiple feature sets, or “views,” to improve performance by capturing complementary insights. Widely applied in fields like image analysis, NLP, and bioinformatics, it enhances generalization performance, robustness to noise, and accuracy by integrating diverse perspectives, often outperforming single-view approaches \cite{41} \cite{42}. Currently, multi-view learning has become a widely accepted solution in mammography analysis \cite{36} \cite{65}. Here, we will not make further elaboration.

\begin{figure*}[h]
    \centering
    \includegraphics[width=0.8\textwidth]{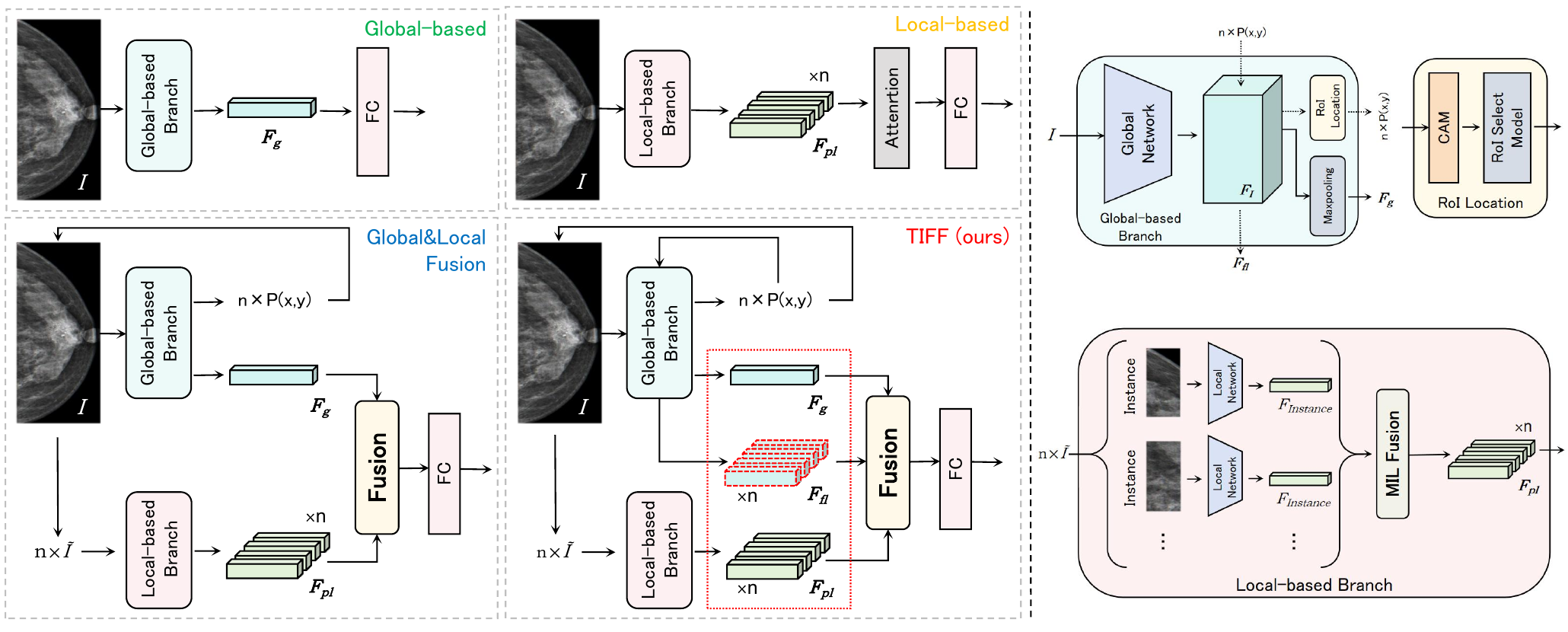}
    \caption{Brief overview of the workflows for the three main existing paradigms and our Tri-level information fusion framework (TIFF). The red dashed box is the main difference.}
    \label{gl}
\end{figure*}

In addition, mammography typically possess extremely high resolution $(3518\times2800)$, with lesions occupying only a very small area. Some lesions can be distributed across a large area, encompassing the entire breast, while others may be confined to a region as small as about $(10\times10)$ pixels. Handling such high-resolution images with lesions of various sizes, poses significant challenges for traditional networks. Currently, there are three mainstream paradigms for handling ultra-high-resolution images. And we illustrate the workflows of these three existing paradigms along with our proposed Three-level Information Fusion Framework (TIFF) workflows in Figure \ref{gl}.

\textbf{Global-based Methods:} Those methods input the entire super-resolution image into the network to capture global information and perform classification tasks \cite{34} \cite{63} \cite{75}.

However, in the context of breast mammography images, the majority of the image area often lacks informative content, and lesions are typically unevenly distributed or discrete. These factors pose significant challenges to classification tasks. Additionally, retaining as much image information as possible for classification necessitates deeper model architectures, which are prone to issues such as vanishing or diminishing gradients, thereby affecting the training process.
Thus, relying solely on global information provided by the image is insufficient for effective early screening of mammography.

\textbf{Local-based Methods:} Pathological image processing faces similar challenges, and its methods often provide us with valuable insights. In pathological image processing, Multiple Instance Learning (MIL), a form of weakly supervised learning Local-based method \cite{79} \cite{77} \cite{76} is commonly used. This approach treats the entire image as a labeled bag, dividing it into multiple instance patches, extracting features each patch, and finally aggregating the features to obtain the final classification of the image (bag).

However, MIL models struggle to accurately pinpoint key instances and identify which instances contribute to the final classification, resulting in noisy learning processes and poor interpretability. Furthermore, in mammography, instances containing lesion areas are scarce, leading to class imbalance that hinders model convergence. 
Consequently, methods which focuses solely on local information, cannot effectively address current limitations. 

\textbf{Global and Local Fusion Methods:}.
Relying solely on global or local information is address the challenges in mammography; therefore, efforts have been made to explore the fusion of both types of information to overcome these difficulties.

\begin{figure*}[h]
    \centering \includegraphics[width=0.8\textwidth, angle=0]{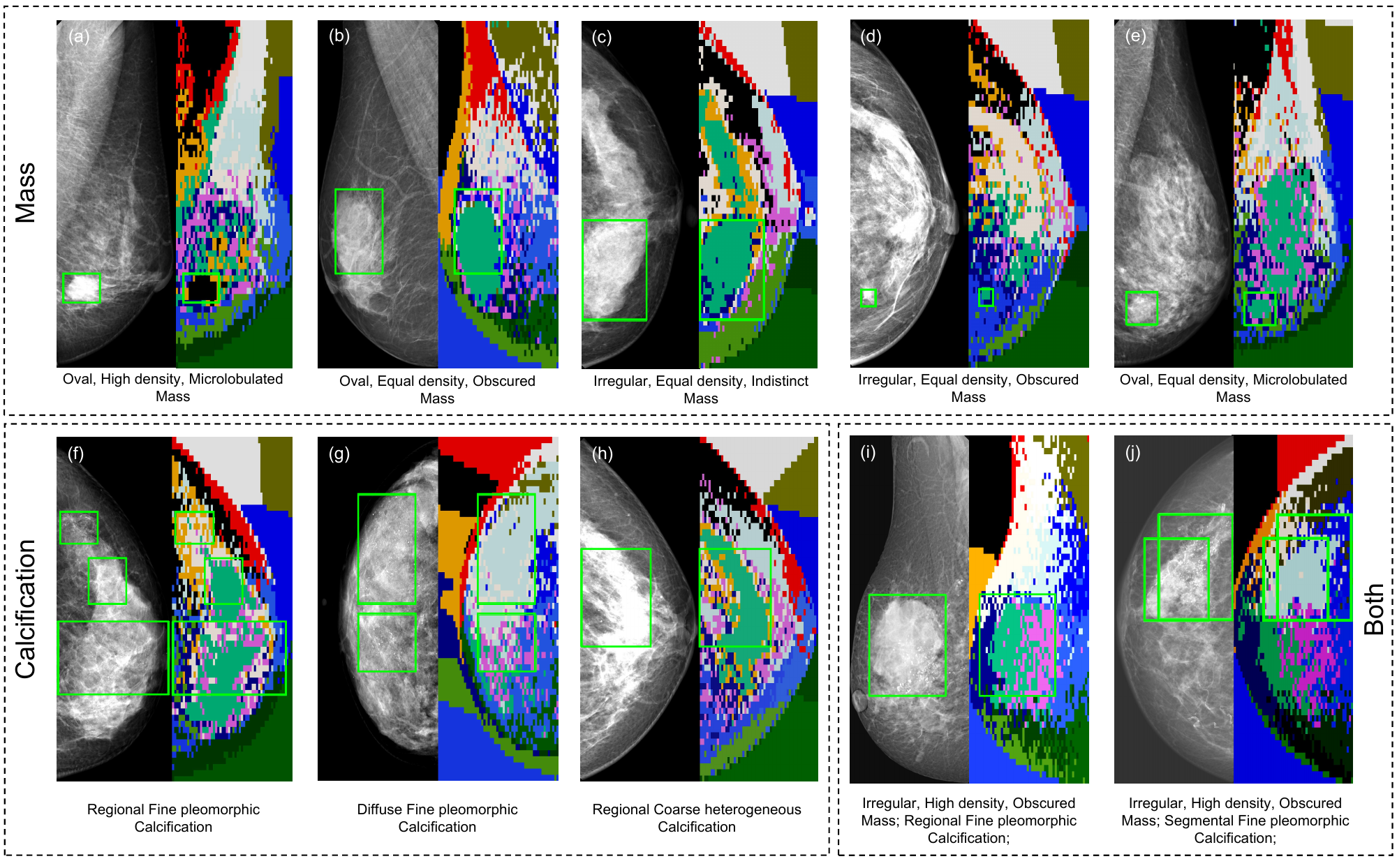}	
    \caption{
    Context Clustering Visualization Diagram. 
    Figures (a) to (j) show that the left half of each image shows the original mammography with annotated suspicious lesions, while the right half presents Contextual clustering visualization, akin to a CNN heatmap and a Attention map, with the suspicious lesion locations also outlined. 
    } 
    \label{fig_1}
\end{figure*}

The Globally-Aware Multiple Instance Classifier (GMIC) framework is a prominent example \cite{25}. GMIC initially uses a global network to extract global information for coarse lesion localization, then refines patch-level images for detailed local information extraction, ultimately combining both types of information for classification. This local information extraction concept is akin to MIL. And the relevant approach introduces multi-view learning based on GMIC, achieving performance improvement by integrating multi-view feature information through pooling \cite{69}. Moreover, the weakly supervised patch-level selection mechanism in GMIC implicitly achieves lesion localization. By considering metrics such as Recall and miss rate, which are more aligned with clinical early screening tasks rather than the general IoU score, the approach achieves promising results, demonstrating its potential to predict reasonable lesion location for clinical early screening.

Although GMIC has shown promising results in early screening tasks for mammography, this cross-scale feature fusion method does not effectively utilize information from each scale. For instance, the global extraction network derives feature information from the entire image, but this information is crudely processed by a max-pooling module before feature fusion, leading to significant wastage of global image features. Additionally, the patch-based local information lacks connection with the corresponding global image information, which remains isolated, patch-based information, thus inevitably results in a fragmented perspective of feature information and isolation between the two feature extraction modules.

\begin{table}[ht]
    \centering
    \caption{Comparison of Model Structures. There A is Multi-view Structure, B means based Global Information, C means based Patch-based Local Information, D means based Feature-based Local Information. }
    \resizebox{0.49\textwidth}{!}{
    \begin{tabular}{l|cccc}
    \hline
    Model & A & B & C & D \\ \hline
    AbMIL\cite{79}          & \xmark & \xmark & \cmark & \xmark \\ \hline
    DsMIL\cite{77}          & \xmark & \xmark & \cmark & \xmark \\ \hline
    TransMIL\cite{76}       & \xmark & \xmark & \cmark & \xmark \\ \hline
    SV Res\cite{63}         & \xmark & \cmark & \xmark & \xmark \\ \hline
    SV SwinT\cite{75}       & \xmark & \cmark & \xmark & \xmark \\ \hline
    GMIC\cite{25}           & \xmark & \cmark & \cmark & \xmark \\ \hline
    MV Res\cite{63}         & \cmark & \cmark & \xmark & \xmark \\ \hline
    MaMVT\cite{65}          & \cmark & \cmark & \xmark & \xmark \\ \hline
    MV GMIC\cite{69}        & \cmark & \cmark & \cmark & \xmark \\ \hline
    Mammo-Clustering(ours)  & \cmark & \cmark & \cmark & \cmark \\ \hline
    \end{tabular}
    }
    \label{table1}
\end{table}

Moreover, both its global and local modules use ResNet for feature extraction, which is not the optimal approach. Traditional feature extraction paradigms, such as convolutional neural networks (CNNs) and attention mechanisms, often face significant challenges when extracting features from entire ultra-high-resolution images. One major issue is the substantial increase in number of pixels, which leads to significant memory and computational resource consumption. Additionally, the limited receptive field of CNNs makes it difficult to effectively capture long-range dependencies in ultra-high-resolution images. More critically, due to the small size of lesion areas, traditional CNNs may lose microcalcifications or fine structures during layer-by-layer downsampling (e.g., pooling), significantly affecting diagnostic accuracy. Attention mechanisms may also overly focus on irrelevant background areas, introducing noise.

\begin{figure*}[h]
    \centering
    \includegraphics[width=1\textwidth, angle=0]{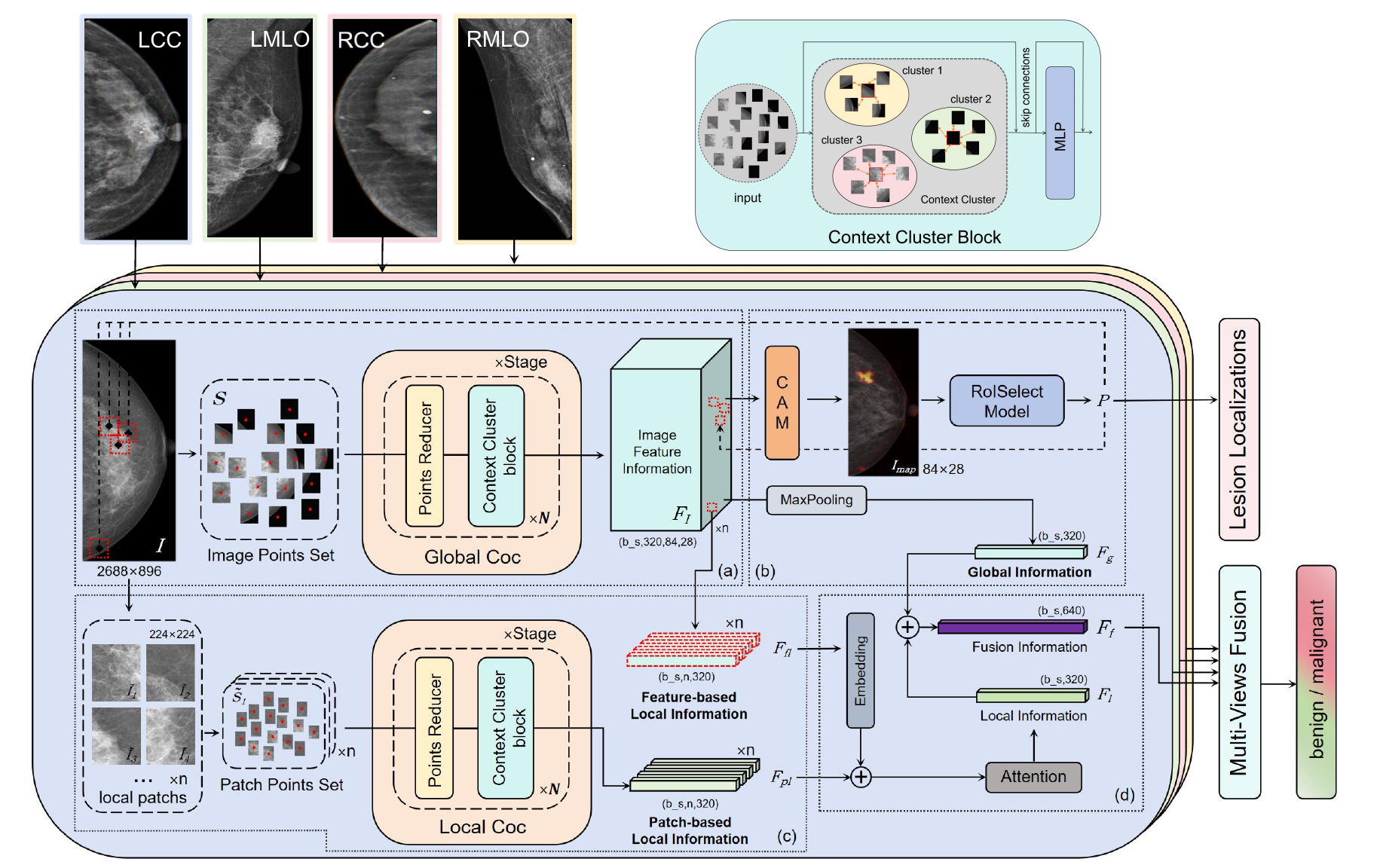}	
    \caption{
    Architecture of the proposed model. $b\_s$ is the abbreviation for batch size.
    } 
    \label{fig_2}
\end{figure*}

\textbf{Tri-level information fusion framework:} Our proposed TIFF categorizes the features of mammography images into Global Information, Feature-based Local Information, and Patch-based Local Information. Feature-based Local Information is derived by selected informative regions, determined by the Saliency Map, using the corresponding Global Information. By deeply integrating this with Patch-based Local Information, it achieves a more comprehensive and robust Local Information, further mitigating the negative impact of the fragmented Patch-based Local Information. We hypothesis that these three levels of information can maximize the utilization of mammography data, the comparisons of different models in the Experiment and Result section also validate our hypothesis. 

To effectively address the above issues, we introduce a Three-level Information Fusion Framework (TIFF) based on a clustering paradigm, Context Clustering (Coc) \cite{74}. The main distinctions of our method are highlighted in Figure \ref{gl} using red dashed box, with the dashed box emphasizing the differences in feature information integration between the TIFF paradigm and other existing paradigms. Additionally, as shown in Table \ref{table1} and Figure \ref{gl}, other methods typically focus either on global or local information, limiting their ability to fully capture relevant features at both levels. In contrast to traditional methods, TIFF achieves more comprehensive information coverage and more effective integration of global and local information.

Finally, to address the high memory and computational costs, limited receptive field, and information loss due to downsampling in traditional Convolutional Neural Networks (CNNs), we adopted the Context Clustering (CoC) method \cite{74}. Compared to conventional CNNs, CoC models images as unordered sets of implicit positional points, thereby mitigating the effects of downsampling and receptive field constraints while requiring fewer computational resources. Moreover, CoC demonstrates superior generalization across diverse data domains, facilitating seamless integration of multi-modal medical data such as mammography, ultrasound, and CT scans.
We demonstrate the performance of this method across various lesion morphologies in Figure \ref{fig_1} and compare this clustering paradigm with CNN and attention paradigms in Figure \ref{cam}, validating its efficacy in mammography. The clustering results are more easily associated with anatomical or pathological features and align with clinical needs. Additionally, common calcified lesions in mammography screening tasks often exhibit clustered or diffuse distributions, naturally aligning with the clustering paradigm.

To the best of our knowledge, Coc is the first method to apply the clustering paradigm to visual representation. Other clustering methods, such as SLIC \cite{66}, are typically used for image preprocessing or specific tasks like semantic segmentation.

We focus on the limitation that the resolution of mammography is too high and lesions occupy only a small portion of the image, and propose  Weakly Supervised Multi-view Tri-level Information Fusion Context Clustering Network. Our primary contributions include:
\begin{itemize}
    \item[$\bullet$] We introduces a novel non-CNN, non-attention-based feature extraction method using \emph{Context clustering} for early breast cancer screening in mammography.
    \item[$\bullet$] We propose a fusion mechanism named Tri-level Information Fusion Framework (TIFF) to integrate global, feature-based local, and patch-based local information, with enhanced focus on local details.
    \item[$\bullet$] Our method achieves state-of-the-art accuracy with the less of parameters among comparable techniques, ensuring efficiency.
\end{itemize}

\section{Method}

\subsection{Overall Framework}

\begin{figure*}[h]
	\centering 
	\includegraphics[width=0.8\textwidth, angle=0]{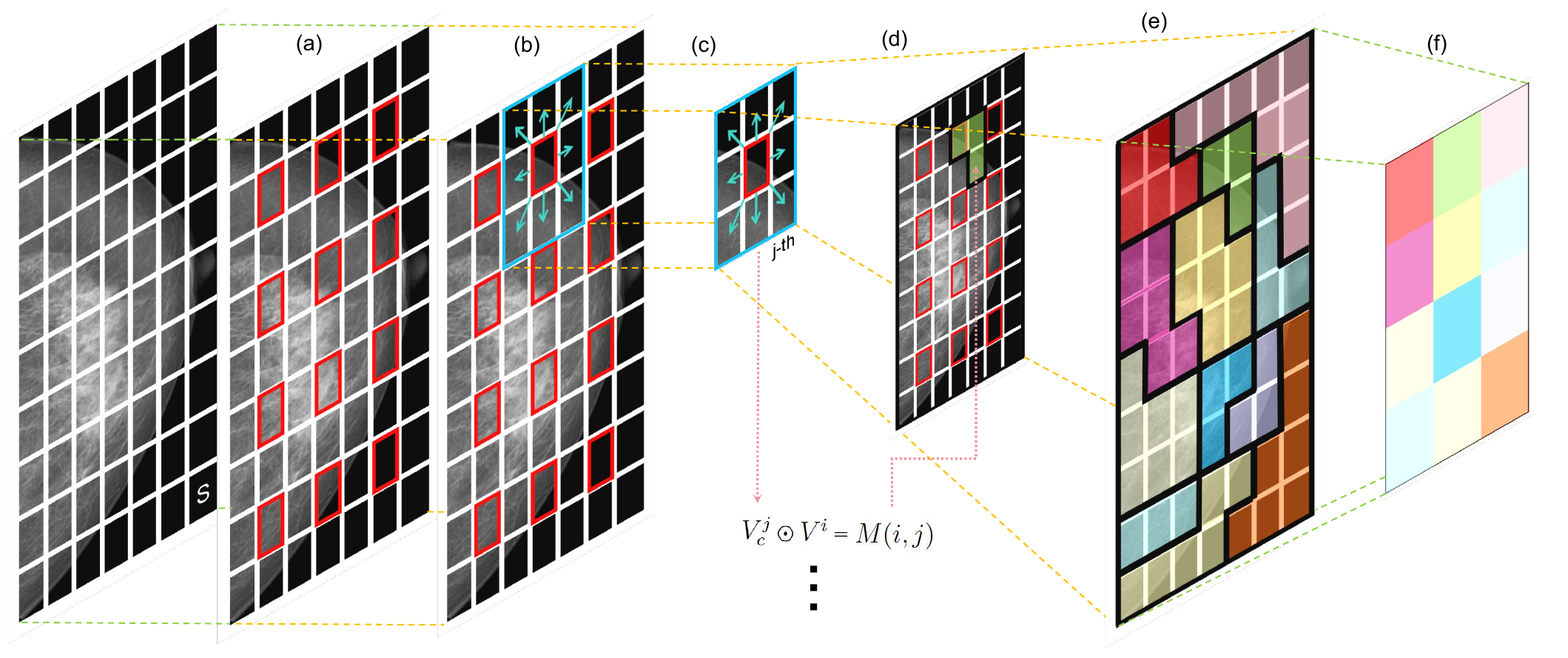}	
	\caption{
        Visual explanation of Context Clustering. Clustering process consists of five components: (a) central anchor points selection; (b) nearest neighbor points search; (c) anchor points central feature computation; (d) similarity analysis; (e) integration of clustering results. (f) point-level downsampling.
        }
	\label{fig_mom1}
\end{figure*}

The proposed Mammo-Clustering model is shown in Figure \ref{fig_2} and can be expressed as follows:

(a) The network input consists of images from four views of the same patient. For each views' image $I$, a point-level enhancement operation is performed, transforming all pixels into a five-dimensional point set composed of color and positional information. This point set is represented as $S \in \mathbb{R}^{5, w \times h}$, where the number of points in the set is $w \times h$.

The set $S$ is then fed into the first Context-Clustering network (Global Coc) to extract Image Feature Information $F_I$.

(b) $F_I$ is processed in two ways: first, through a feedforward network (CAM) to obtain a Saliency Map, denoted as $I_{map}$; second, through a Maxpooling module to obtain global information $F_{g}$ for later use.

Based on the Saliency Map, the RoI selection module outputs a set of position information $P$, which are represented as:
\[
P = \{ p_{1}, p_{2}, \ldots, p_{n}\}
\]
$p_{n}$ $(x_{n}, y_{n})$ denotes the coordinates of the top-left corner of the n-th selected patch-level RoI, each of which defines a patch-level image with height $h_{\tilde{I}}$ and width $w_{\tilde{I}}$. The number of selected RoIs \text{n}, as well as the height $h_{\tilde{I}}$ and width $w_{\tilde{I}}$ of each patch-level image, can be manually specified as needed. In this work, they are set to 4, 224, and 224, respectively. RoI selection is independent of dataset annotations.

(c) With $P$, we locate \text{n} patches $\tilde{I}$ on the original image $I$ circled by the red dotted line in Figure \ref{fig_2} and obtain \text{n} Feature-based Local Information $F_{fl}$ from Image Feature Information $F_{I}$. 

Each selected patch $\tilde{I}$ also undergoes point-level enhancement to obtain $\tilde{S}$, which is fed into the second Context-Clustering network (Local Coc) to extract Patch-based Local Information $F_{pl}$.

(d) Subsequently, Local Information $F_{l}$ is obtained through an attention mechanism module $f_{atten}$ by fusing Feature-based Local Information $F_{fl}$ and Patch-based Local Information $F_{pl}$.

\begin{equation}
    F_l = f_{atten}(F_{fl}\oplus F_{pl})
\end{equation}
The $\oplus$ represents the fusion operation of the two types of information, which is then processed through an attention mechanism to enhance relevant features information. The local information $F_l$ is then fused with the previously mentioned global information $F_g$ to generate multi-instance fusion information $F_f$ for the current view.

Through the above four process, we obtain the corresponding $F_f$ for images from four views (bilateral craniocaudal (CC) and mediolateral oblique (MLO) views of both breasts of the same patient), which are integrated into the final Multi-views fusion information for the binary classification of benign and malignant cases.

\subsection{Data Preprocessing}

Mammography data are often stored in Dicom format, but networks typically cannot directly read Dicom data. Additionally, the high resolution of mammography images poses challenges to the GPU memory size used during network training. To address these two major difficulties, we need to first convert Dicom data to PNG format and remove redundant, unnecessary parts from the mammography images.

Once the image is obtained in PNG format, it is necessary to crop the large black unused areas. The image, converted from DICOM to PNG, is first read as a grayscale image. Edge detection is performed using the Canny algorithm \cite{82} to identify edges within the image. Dilation and erosion operations are then applied to the edge-detected image to form clear boundaries. Contours within the image are identified, and the largest contour is selected as the target area. The image is cropped based on the contour boundaries to remove unused regions. During cropping, consideration is given to the breast’s left or right position $(breast\_side)$ to ensure precision. Finally, the cropped image is saved as a 16-bit depth PNG file.

\subsection{Network Structure Details}

\subsubsection{From Image to Set of Points}
The point-level enhancement operation for images involves transforming each pixel of an image with dimensions (3, h, w) into a five-dimensional information composed of RGB information and positional information, with dimensions (r, g, b, x, y). This method abstracts the image into a point set.

This approach allows us to perform feature extraction through simple clustering. From a global perspective, the image is treated as a collection of unordered discrete data points with color and positional information. Through clustering, all points are grouped into clusters, each containing a centroid. Since each point in the set includes color and positional information, the cluster implicitly contains spatial and image information.

\subsubsection{Context Clustering}
Context Clustering module can be divided into two parts: Feature Aggregating and Feature Dispatching. It processes the point set $S \in \mathbb{R}^{5, w \times h}$, enhanced at the point level, using context clustering blocks within a multi-layer structure to extract multi-scale image feature information. A Points Reducer Block is linked before each layer’s context clustering block to enhance computational efficiency by reducing the number of points. The context clustering process can be divided into five core stages: (a) central anchor point selection; (b) nearest neighbor points search; (c) anchor points central feature computation; (d) similarity analysis; (e) integration of clustering results. (f) point-level downsampling. We also illustrate this clustering process in Figure \ref{fig_mom1}.

\begin{enumerate}[label=\alph*.]
    \item Evenly select some central anchor points from the point set $S$ in space.
    
    \item Central anchor points and their nearest neighbor points are connected and fused through linear projection. The number of neighboring points, $k$, can be customized. If all points are ordered sequentially and $k$ is set to 4 or 9, this downsampling can be achieved via convolution operations. Figure \ref{fig_mom1} illustrates the case where $k$ is 9.
    
    \item Compute the central feature value $V_{c}$ for each central anchor point by averaging the feature values $V_{n}$ of its $k$ nearest neighbor points.
    \[
    V_{c} = \frac{\sum_{i=1}^{k} V_{n}^{i}}{k} 
    \]
    where $V_{n}$ is determined by the RGB and positional information of the point.

    \item Calculate the cosine similarity matrix $M$ between the point $s$ in $S$ and the set of central points. Assign the point to the most similar center based on the similarity matrix.

    \[
    M(i, j) = V^i \odot V^j_c = \frac{V^i \cdot V^j_c}{\|V^i\| \|V^j_c\|}
    \]
    where $\odot$ is the cosine similarity calculation method, $V^i$ represents the feature values of the i-th point in $S$, and $V^j_c$ represents the feature values of the j-th central anchor point.
    
    \item The clustering method follows the traditional SuperPixel \cite{73} approach SLIC \cite{66}, assigning each point to the most similar center, resulting in $c$ clusters. Each cluster may contain a different number of points.

    \item Subsequently, the clustered centroids are retained to perform point-level downsampling, allowing step (a) to be continued in the next layer. Both steps (a) and (f) are executed by the Points Reducer module. In the first or last layer, steps (a) and (f) are performed selectively.
\end{enumerate}

The point set $S$ consists of the RGB information and positional data of each pixel in the image. Therefore, the features extracted through this clustering paradigm inherently encapsulate the correlation between image and spatial information \cite{74}.

In the Feature Dispatching section, the aggregated features within each cluster are adaptively assigned to each point based on similarity. Points can communicate with each other and share features from all points within the cluster.

\subsubsection{Image to Patch Selection}
We need to select \text{n} suspicious patches $\tilde{I}$ from the original image $I$ from a global perspective, which will provide us with feature information from a local perspective. This selection process relies on the collaboration between the CAM module and the RoI Select Model.

The Image Feature Information $F_{I}$ extracted by Global Coc is fed into the feedforward network CAM module to generate a $h_{map} \times w_{map}$ Saliency Map for image feature visualization, where $h_{map}$ and $w_{map}$ are manually set. The CAM module consists of a 2D convolutional layer with a kernel size of 1 and retains gradients for iterative optimization during training \cite{81}.
The Saliency Map is normalized and fed into the RoI Select Model for greedy search of regions of interest. In each iteration, the algorithm greedily identifies each region and selects the top \text{n} regions with the highest total weights among all current regions, with weights determined by average pooling. The coordinates of all regions are added to the set of position information $P$. During the selection process, a mask is applied to the selected regions to prevent redundant selection.

The coordinates of these selected regions are ultimately mapped to the original image size to obtain n patch-level images of size $h_{\tilde{I}} \times w_{\tilde{I}}$. The heights and widths of the three types of images, including $h_{\tilde{I}}$, $w_{\tilde{I}}$, $h_{map}$, $w_{map}$, $h_{I}$ and $w_{I}$ are manually set and required to ensure that the height and width of the original image $I$ can always be divisible by those of the other two images.

Figure \ref{fig_mom2} visualizes the selected patch-level images within the original image and compares the locations of these patch-level images on the source image with the suspicious lesion locations outlined by the physician.

\subsubsection{Tri-level Information Fusion}
Tri-level Information primarily refers to three types of feature information with different sizes and focuses: global information, Feature-based Local Information, and Patch-based Local Information.

\textbf{Global Information:} $F_g$, with size $(batch \:size, \: dim)$, is obtained by applying a Maxpooling module to the Image Feature Information $F_I$ extracted by the Global Coc. $F_g$ represents the global feature information of the image from a macroscopic perspective. where $batch\:size$ is the number of samples used in each iteration of training and $dim$  represents the number of feature channels of the image point set after downsampling through multiple levels of the Context Clustering Module.

\textbf{Feature-based Local Information:} $F_{fl}$ is derived from $F_I$, similar to $F_g$. However, $F_{fl}$ is obtained by selecting and extracting information from $F_I$ based on \text{n} patch-level images selected by the model. The size of $F_{fl}$ is $(batch\:size, \: n,\: dim)$. $F_{fl}$ emphasizes the local representation of global feature information within the selected regions of interest.

\textbf{Patch-based Local Information:} $F_{pl}$ is obtained by the Local Coc through Context Clustering feature extraction on n patch-level images selected by the model. The size of $F_{pl}$ are consistent with those of $F_{fl}$. $F_{pl}$ represents the local feature information of the chosen regions of interest.

This approach yields feature representations of the original image and patch-level images, encompassing both local and global perspectives. The complementarity between these types of information maximizes the utilization of mammography data features, providing enhanced support for the model.

\textbf{Information Fusion:} $F_{fl}$ is aligned with $F_{pl}$ through a trainable Embedding Module and concatenated to obtain feature information. The feature information's size is $(batch\:size,\: 2n,\: dim)$. The Embedding Module consists of an MLP. This feature information is then fed into an attention mechanism module to obtain Local Information $F_{l}$, which has the same size as $F_{g}$. The use of the attention mechanism serves two purposes: firstly, it facilitates the fusion of different features and the learning of interrelated information; secondly, it mitigates potential Patch-level image redundancy, as not all of the n selected Patch-level images necessarily carry beneficial information, which could impact training.

Finally, $F_{g}$ is concatenated with $F_{l}$ in $dim$ dimension to form the final Fusion Information $F_{f}$ from the current perspective. The $F_{f}$ from different views are first merged together and then fused through an attention mechanism to achieve Multi-view integration. Similar to the processing of $F_{fl}$ and $F_{pl}$, the use of the attention mechanism not only allows for the fusion of $F_{f}$ from different perspectives and the learning of interrelated information but also mitigates the impact of potential view-level image redundancy on training, as not all views images carry useful information in most cases.

It is noteworthy that not only $F_{f}$ is processed by the attention module; $F_{g}$ and $F_{l}$ are also individually processed by the attention mechanism. This is because we aim to optimize different components of the network structure based on the loss obtained from various features.

\subsection{Model Train Details}

\subsubsection{Implementation Details}
In this study, we evaluated the breast cancer early screening task on two public datasets using various approaches, including MIL, Single-view, and Multi-view methods, and compared them with our proposed Tri-level Information Fusion Context Clustering Framework. All experiments were conducted on a single NVIDIA 3090 24G GPU, using Adam as the optimizer \cite{80}. A fixed-step learning rate (StepLR) decay strategy was employed to fine-tune the learning rate, preventing overfitting and ensuring better convergence to the optimal solution.

\subsubsection{Loss Function}
We chose a composite loss function to achieve targeted optimization of different components.
\[
L = \alpha \cdot L_{g} + \beta \cdot L_{l} + \gamma \cdot L_{f} + \delta \cdot L_{map}
\]
This composite loss function consists of three losses, $L_{g}$, $L_{l}$ and $L_{f}$, which are the losses between the predicted and true values based on different features, along with $L_{map}$ the weighted average intensity of a Saliency Map under the L1 norm.

The BCEWithLogitsLoss function is used for $L_{g}$, $L_{l}$ and $L_{f}$. The weights $\alpha$, $\beta$, $\gamma$, and $\delta$ represent the proportional influence of each loss, and they are set independently. Here, we consider $L_{g}$, $L_{l}$ and $L_{f}$ to be equally important, so we recommend setting $\alpha$, $\beta$ and $\gamma$ to 1. Since the value of $L_{map}$ is often relatively large, we suggest setting $\delta$ to a value less than 0.001.

\section{Experiment and Result}

\subsection{Datasets}

\subsubsection{Vindr-Mammo}
The Vindr-Mammo \cite{67} dataset is a large-scale annotated collection of full-view digital mammography images, consisting of full-view examination images from 4,999 volunteers (officially claimed to be 5,000, but we identified one erroneous data entry). Each case is meticulously annotated with clinical indicators such as lesion type, breast density level, BI-RADS category, and lesion location. Each case is independently reviewed, with disagreements resolved by a third radiologist through arbitration. The dataset is extensive, with high-quality images and detailed annotations for clinical downstream tasks; however, it exhibits an imbalance in the distribution of benign and malignant cases, leading to a long-tail issue.

\subsubsection{CBIS-DDSM}
The CBIS-DDSM dataset (Curated Breast Imaging Subset of the Digital Database for Screening Mammography) \cite{68} is a widely utilized resource in breast cancer research. It comprises 1,645 digitized mammographic images with detailed annotations regarding lesion ROI, lesion type (e.g., calcification, mass), breast density, and BI-RADS categories. Importantly, the labels (benign or malignant) are pathologically confirmed. Its comprehensiveness and balanced data distribution make it a standard benchmark for evaluating AI model performance in mammography-based studies. However, the data was collected in earlier years, with limitations in both its quantity and quality.

\begin{table}[h]
\caption{The composition of data for the two datasets.}
\renewcommand{\arraystretch}{1.2}
\resizebox{0.5\textwidth}{!}{
\begin{tabular}{l c c c} 
\hline
 & \multicolumn{3}{c}{\textbf{Vindr-Mammo}} \\
 \hline
    & Benign & Malignant & Total \\ 
 \hline
    Training & 3,614(90.37\%) & 385(7.63\%) & 3,999  \\ 
    Test     & 904(90.40\%)   & 96(9.60\%)  & 1,000  \\ 
    Overall  & 4,518(90.38\%) & 481(9.62\%) & 4,999  \\ 
 \hline
 & \multicolumn{3}{c}{\textbf{CBIS-DDSM}} \\
 \hline
    & Benign & Malignant & Total \\ 
 \hline
    Training & 684(52.90\%)  & 609(47.10\%) & 1,293 \\ 
    Test     & 203(57.67\%)  & 149(42.33\%) & 352   \\ 
    Overall  & 887(53.92\%)  & 758(46.08\%) & 1,645 \\ 
    \hline
\end{tabular}
}
\label{Table1}
\end{table}

Both Vindr and CBIS-DDSM provide detailed annotations of lesion locations, but such annotation tasks are generally high-cost. By employing weakly supervised learning to enable the network to autonomously localize lesion positions, the cost of dataset creation can be significantly reduced.

\subsection{Performance Indicator}

\begin{table*}[h]
\centering
\caption{Performance of each model on two datasets. SV and MV represent Single-View and Multi-Views, respectively. Backbone of the network after "-".}
\renewcommand{\arraystretch}{1.2}
\resizebox{1\textwidth}{!}{
\begin{tabular}{lccccccc}
\hline
    & \multicolumn{3}{c}{\textbf{Vindr-Mammo}} & \multicolumn{3}{c}{\textbf{CBIS-DDSM}} \\
    \hline
    &AUC & ACC & F1 score &AUC & ACC & F1 score & Params(M) \\
    \hline
    AbMIL\cite{79} $_{\text{ICML'18}}$        & $0.618 \pm 0.02$ & 0.776 & 0.825 & $0.726 \pm 0.02$ & 0.571 & 0.671 & 0.4   \\
    DsMIL\cite{77} $_{\text{CVPR'21}}$        & $0.605 \pm 0.02$ & 0.730 & 0.781 & $0.697 \pm 0.02$ & 0.500 & 0.583 & \textbf{0.2} \\
    TransMIL\cite{76} $_{\text{NeurIPS'21}}$   & $0.631 \pm 0.02$ & 0.888 & 0.890 & $0.739 \pm 0.02$ & 0.635 & 0.637 & 2.5  \\
    SV Res18\cite{63} $_{\text{TMI'20}}$       & $0.727 \pm 0.02$ & 0.783 & 0.821 & $0.719 \pm 0.02$ & 0.646 & 0.639 &  1.4  \\
    SV SwinT\cite{75} $_{\text{ICCV'21}}$      & $0.731 \pm 0.02$ & 0.651 & 0.719 & $0.724 \pm 0.02$ & 0.651 & 0.599 & 14.1 \\
    GMIC-Res18\cite{25} $_{\text{MIA'23}}$     & $0.793 \pm 0.02$ & 0.847 & 0.878 & $0.778 \pm 0.02$ & 0.682 & 0.680 & 22.4 \\
    MV Res18\cite{63} $_{\text{TMI'20}}$       & $0.740 \pm 0.02$ & 0.753 & 0.796 & $0.731 \pm 0.02$ & 0.676 & 0.641 & 6.1  \\
    MaMVT\cite{65} $_{\text{MIA'24}}$          & $0.770 \pm 0.02$ & 0.882 & 0.867 & $0.749 \pm 0.02$ & 0.649 & 0.649 & 30.7 \\
    MV GMIC-Res18\cite{69}             & $0.797 \pm 0.02$ & 0.887 & 0.879 & $0.781 \pm 0.02$ & 0.699 & 0.691 & 22.6 \\
    MV GMIC-SwinT                       & $0.799 \pm 0.02$ & 0.874 & 0.854 & $0.785 \pm 0.02$ & 0.694 & 0.694 & 28.8 \\
    Mammo-Clustering(ours)               & $\textbf{0.828} \pm 0.02$ & \textbf{0.919} & \textbf{0.906} & $\textbf{0.805} \pm 0.02$ & \textbf{0.709} & \textbf{0.709} & 9.8 \\ \hline
\end{tabular}
}
\label{Table2}
\end{table*}

We analyze the performance of our breast cancer early screening model from two perspectives: first, the classification indicators, which assess the final benign or malignant classification performance; second, the localization indicators, which evaluate the model’s ability to locate lesion areas through patch-level image selection under unsupervised conditions.

\subsubsection{Classification indicators}
\begin{itemize}
    \item AUC (Area Under the Curve): AUC means the area under the receiver operating characteristic (ROC) curve. The ROC curve uses the true positive rate for mammography benign-malignant classification as the y-axis and the false positive rate as the x-axis. It provides an aggregate measure of performance across all possible classification thresholds. A higher AUC value indicates a better model performance, with 1 representing a perfect model and 0.5 a random guess.
    
    \item ACC (Accuracy): Accuracy is the proportion of true results (both true positives and true negatives) among the total number of cases examined, the corresponding clinical term is “specificity”. It gives a straightforward measure of how often the classifier is correct.

    \item F1 Score: The F1 Score is the weighted average of Precision and Recall. This score takes both false positives and false negatives into account. Given the long-tail issue in the data, we selected the micro F1 score as the evaluation metric, it is particularly useful when the class distribution is uneven. 
    
\end{itemize}

\subsubsection{Localization indicators}
\begin{itemize}
    \item MDR(Miss Detection Rate):
    MDR is defined as the percentage of the number of undetected suspicious lesion areas $N_{miss}$ relative to the total number of suspicious lesion areas $N_{gt}$. Because, in clinical practice, we are more concerned about missed lesions, i.e., false negatives, rather than false positives.

    \item Recall:
    Recall is a metric used in object detection to evaluate a model’s ability to identify all relevant objects in an image. It measures the proportion of actual positive instances (i.e., objects that should be detected) correctly identified by the model. In this context, it reflects the model’s capability to detect all existing lesions.
\end{itemize}

These metrics collectively provide a comprehensive evaluation of the performance of breast cancer screening models, helping to understand their strengths and weaknesses in various aspects of classification and Localization.

\subsection{Comparative Experiment}

In this study, we evaluated several models on two datasets: Vindr-Mammo and CBIS-DDSM. The performance metrics considered were AUC, ACC, and F1 score and so on.

\subsubsection{Assessment of Classfication}
\textbf{Classfication Accuracy:}
For the Vindr-Mammo dataset, Mammo-Clustering (ours) model achieves the highest AUC $0.828 \pm 0.02$, ACC of $0.919$, and F1 score of $0.906$ among all models, achieving approximately a 3.5\% improvement in AUC accuracy compared to the suboptimal model MV GMIC-SwinT, indicating superior performance.

On the CBIS-DDSM dataset, Mammo-Clustering (ours) model again demonstrates the best performance with an AUC of $0.805 \pm 0.02$, ACC of $0.709$, and F1 score of $0.709$, achieving approximately a 2.4\% improvement in AUC accuracy compared to the suboptimal model MV GMIC-SwinT.

Comparing the Single-View model and Multi-Views model reveals that integrating multi-view information can enhance classification accuracy. However, the AUC comparison between GMIC-Res18 and MV Res18 indicates that the fusion mechanism of global information and Local Information in the GMIC architecture is significantly more effective than the integration of features from different angles.

Additionally, We found that the MIL paradigm, commonly used in pathological image classification, did not perform well on the Vindr-Mammo dataset but showed good results on the CBIS-DDSM dataset. This discrepancy may be attributed to the long-tail distribution issue in the Vindr-Mammo dataset.

In this this two datasets, the advantages of Context Clustering and Tri-level Information Fusion architecture are more pronounced than other method, showing significant advantages in AUC.

\textbf{Model Complexity:}
In terms of model complexity, measured by the number of parameters, the ours model had 98.05 million parameters. Other smaller networks often cannot achieve the accuracy of our model and show a significant gap. This is efficient compared to the MaMVT with 30.73 million parameters and the MV GMIC-Res18 with 22.68 million parameters, considering the performance gains achieved.

\begin{figure}[h]
\centering 	
\includegraphics[width=0.5\textwidth, angle=0]{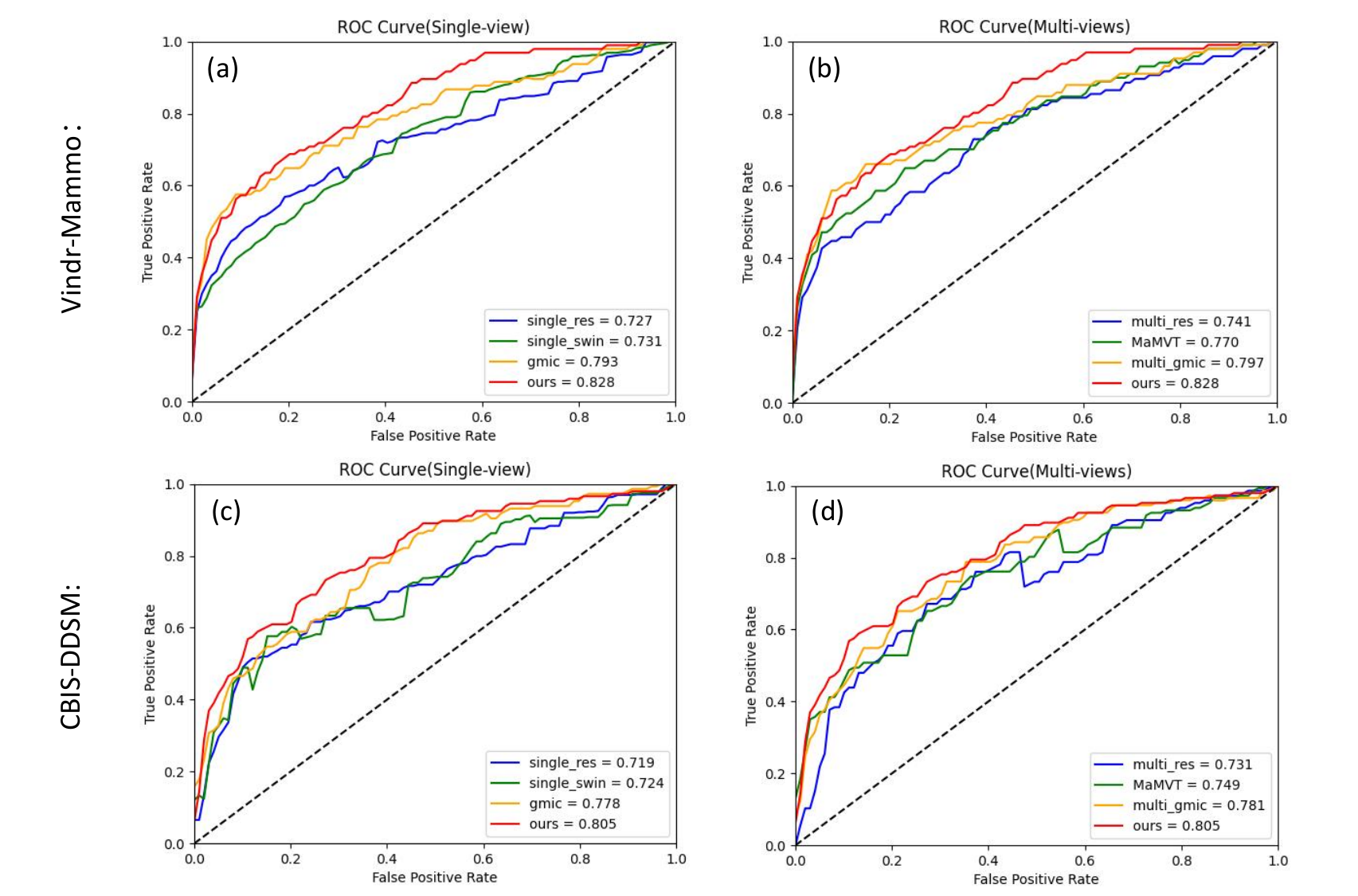}	
	\caption{
    Comparison of ROC curves of different models on two public datasets. Figures a and b compare the ROC curves of our model with other Single-view and Multi-views architectures on the Vindr-mammo dataset. Figures c and d present the ROC curves comparison on the CBIS-DDSM dataset.
} 
	\label{fig_mom3}
\end{figure}

\textbf{ROC curve:}
The ROC curve in figure \ref{fig_mom3} provides insights that cannot be obtained from tables alone. 

Analyzing the ROC curve, we observe that most models, except ours, exhibit a concave shape in the middle. This is due to class imbalance in the data, further validating the effectiveness of our model’s architecture.

Overall, our model offers a robust and efficient approach, achieving state-of-the-art performance on both datasets, surpassing the second-best AUC by over 0.02, with fewer parameters. The global information and Local Information fusion proves effective for both multi-view and single-view models and the multi-view learning approach enhances model performance. It is evident that superior local information will inevitably lead to improved classification performance.

\subsubsection{Assessment of localization}

\begin{figure*}
	\centering 
	\includegraphics[width=1\textwidth, angle=0]{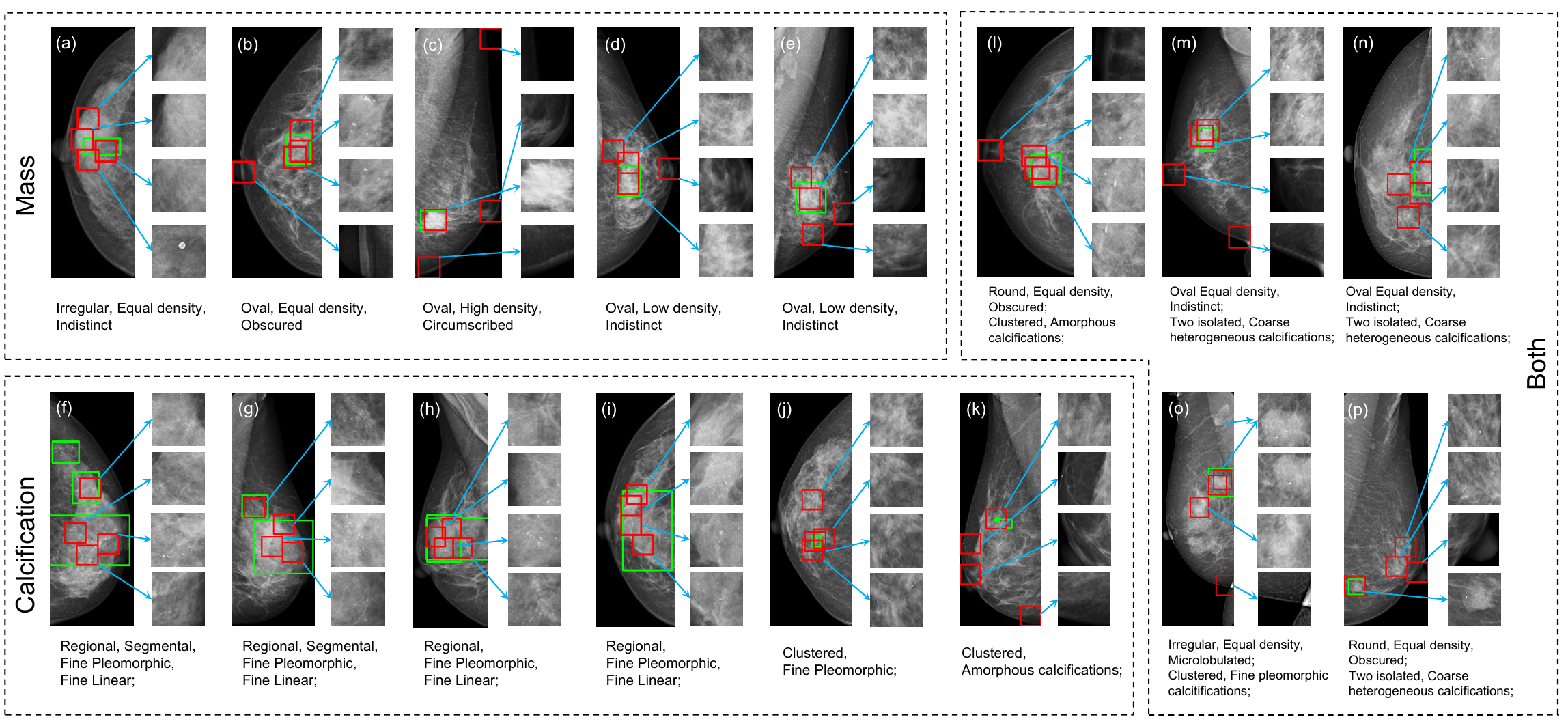}	
	\caption{
        Visualization of Patch-level images extracted by the model. The green box on the mammography indicates the location of the lesion, while the red box represents the Patch-level images selected by the model. 
        } 
	\label{fig_mom2}%
\end{figure*}

We can observe a high degree of overlap between these two boxes in figure \ref{fig_mom2}. Obviously, this visualization demonstrates the great potential of this weakly supervised lesion localization approach. Additionally, the weakly supervised lesion localization performance of our model was evaluated and endorsed by clinicians, who noted that the model could identify some suspicious areas not marked in the dataset. This further validates the model’s generalization and robustness, which will be discussed in detail in the Discussion section.

We compared the performance of three weakly supervised lesion localization models without comparing them to fully supervised models. Our aim was to validate the feasibility of training weakly supervised models on datasets without annotated lesion regions, which would significantly reduce the burden of creating mammography datasets.

\begin{table}[!ht]
    \centering
    \caption{
    Comparison of Different Weakly Supervised Models in Lesion Localization Tasks.
    }
    \renewcommand{\arraystretch}{1.2}
    \resizebox{0.5\textwidth}{!}{
    \begin{tabular}{l c c c}
    \hline
        ~             & MDR   & Recall & AUC  \\ \hline
        GMIC-Res18    & 0.433 & 0.476  & $0.793 \pm 0.02$ \\ 
        MV GMIC-Res18 & 0.336 & 0.643  & $0.797 \pm 0.02$ \\ 
        MV GMIC-Res34 & 0.412 & 0.479  & $0.789 \pm 0.02$ \\ 
        MV GMIC-SwinT & 0.322 & 0.650  & $0.799 \pm 0.02$ \\ 
        \textbf{Mammo-Clustering(ours)}  & \textbf{0.294} & \textbf{0.685} & $\textbf{0.828} \pm 0.02$ \\ \hline
    \end{tabular}
    }
    \label{Table5}
\end{table}

From Table \ref{Table5}, we observe that our model achieves the lowest missed detection rate (MDR) of 0.294 and the lowest recall rate of 0.476, demonstrating its potential. All the comparative models employ the same weakly supervised lesion localization approach as our model, enabling lesion location identification using only classification labels, making them suitable for comparison. We selected four different models encompassing both CNN and attention mechanism paradigms and conducted evaluation analysis on two metrics: MDR and recall. Compared to the second-best model, our approach shows an improvement of approximately 0.03 in both MDR and recall.

\subsubsection{Visual comparison}

\begin{figure*}[ht]
    \centering 
    \includegraphics[width=1\textwidth, angle=0]{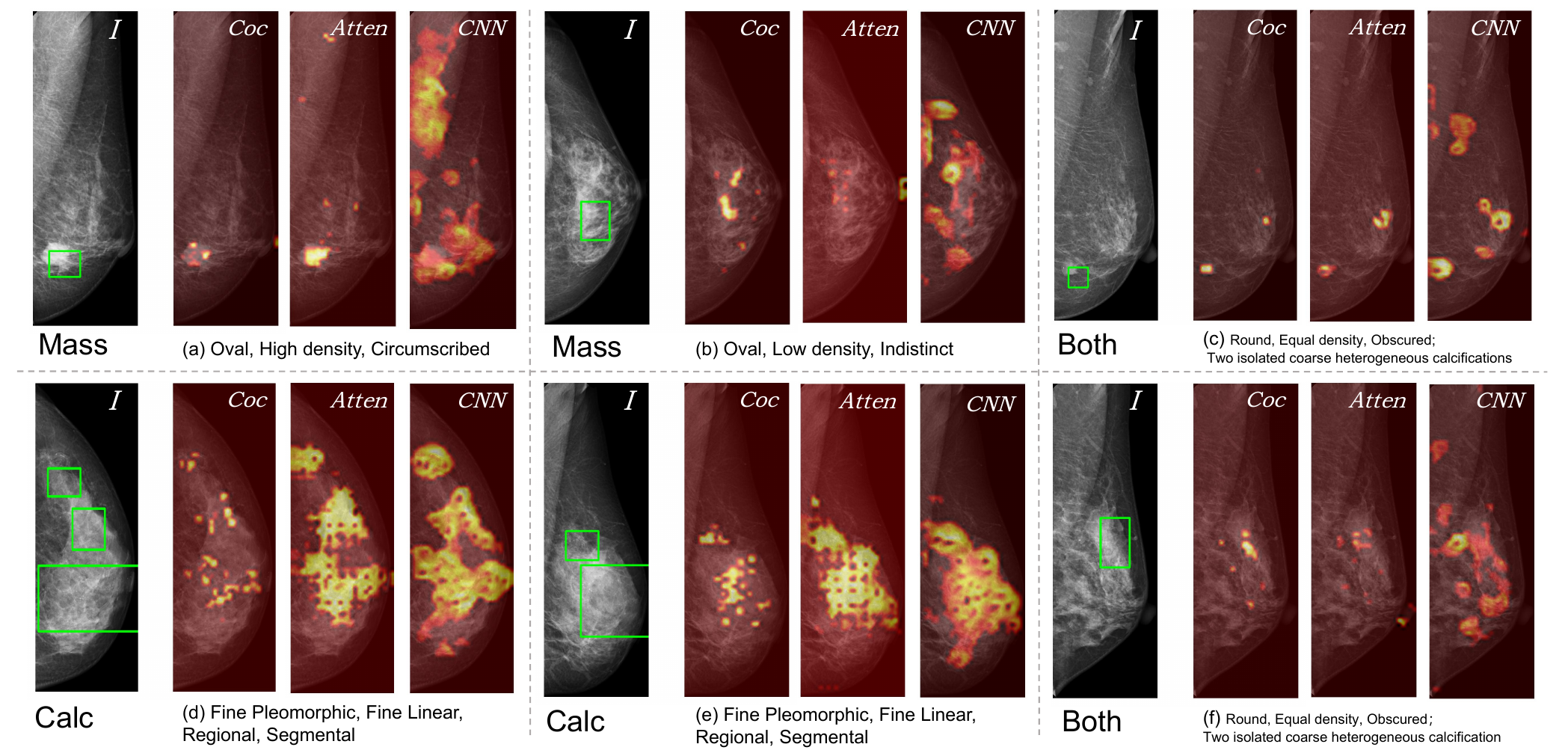}	
    \caption{Comparison of activation map visualizations from different feature extraction paradigms. The green boxes on the original images indicate the annotated true locations of the lesions.
    } 
    \label{cam}
\end{figure*}

We visualized the feature activation maps of different feature paradigms under various lesion types, as shown in Figure \ref{cam}.

In this comparison, we selected representative models, ResNet and SwinT, as exemplars of CNN and Attention mechanisms, respectively, to compare with the Context Clustering paradigm. It is evident that the activation maps based on Context Clustering are clearer and significantly outperform the other two paradigms. According to analysis by professional clinicians, these maps often better align with the actual lesion distribution. Additionally, the activation maps based on Context Clustering demonstrate relatively stable performance across both types of lesions. The activation maps based on CNN perform the worst; although they sometimes identify the location of regions of interest, they exhibit low contrast. Both the attention mechanism-based activation maps and those based on Context Clustering show relatively clear and accurate localization of regions of interest. However, the attention mechanism paradigm often performs poorly on small lesions, possibly due to the feature grid characteristics inherent in the SwinT approach.

\subsection{Ablation Experiment}

\subsubsection{Different Feature Extraction Paradigm}

This ablation study aims to demonstrate the superiority of Context Clustering in feature extraction performance for Mammography through numerical analysis.

\begin{table}[ht]
\centering
\caption{
Performance of different feature extraction paradigm on the Vindr-Mammo, $SV$ represents a single-view learning approach.
}
\renewcommand{\arraystretch}{1.2}
\resizebox{0.5\textwidth}{!}{
\begin{tabular}{l c c c} 
 \hline
                     & AUC  & ACC & F1 score \\ 
 \hline
 SV Res18(CNN-base)        & $0.727 \pm 0.02$ & 0.783 & 0.821 \\
 SV SwinT(Atten-base)      & $0.731 \pm 0.02$ & 0.651 & 0.719 \\
 SV Coc(Clustering-base)   & $\textbf{0.762} \pm 0.02$ & \textbf{0.794} & \textbf{0.833} \\
 \hline
\end{tabular}
}
\label{Table3}
\end{table}

The table clearly demonstrates the superiority of the Context Clustering architecture, achieving the highest AUC as well as the best ACC and F1 scores in single-view learning, indicating its balance in mammography tasks.

\subsubsection{Different Information Fusion Method}

We identified two distinct sources of local information: patch-based local information and feature-based local information. Moreover, this feature-based local information has been overlooked in existing work.

The aim of this ablation study is to validate the effectiveness of our Tri level Information Fusion mechanism that combines these two types of local information with global information.

\begin{table}[ht]
\centering
\caption{
Performance of different information fusion method on the Vindr-Mammo
}
\renewcommand{\arraystretch}{1.2}
\resizebox{0.5\textwidth}{!}{
\begin{tabular}{l c c c} 
 \hline
    & AUC  & ACC & F1 score \\ 
 \hline
 $F_{g}$-based   & $0.783 \pm 0.02$ & 0.815 & 0.852 \\
 $F_{g}$ \& $F_{pl}$ fusion & $0.810 \pm 0.02$ &0.890& 0.891 \\
 $F_{g}$ \& $F_{fl}$ fusion & $0.806 \pm 0.02$ & 0.895 & 0.868 \\
 $\textit{TIFF(ours)}$ & $\textbf{0.828}\pm 0.02$&\textbf{0.919}& \textbf{0.906}\\ 
 \hline
\end{tabular}
}
\label{Table4}
\end{table}

We can clearly observe that integrating only one type of local information with global information does not yield the best results, yet it is significantly better than focusing solely on global information. Adding Patch-based Local Information to global information increases the AUC to 0.810, while adding Feature-based Local Information raises the AUC to 0.806. However, the Tri-level Information Fusion mechanism, which combines all three types of information, achieves the best result with an AUC of 0.828.

\section{Discussion}

\textbf{Performance variation due to differences in dataset distribution:} Although our model achieves state-of-the-art performance on two public datasets, the varying performance of MIL methods across different datasets has drawn our attention. We believe the poor performance of MIL methods on the Vindr-Mammo dataset is due to its long-tail distribution. In this distribution, the model may overfit to the dominant classes, capturing noise rather than meaningful patterns from minority classes. This ultimately leads to biased predictions towards classes or features that are overrepresented in the dataset, resulting in poor generalization to underrepresented classes. Conversely, in the clinical task of early breast cancer screening using mammography, it is crucial to pay extra attention to the underrepresented suspicious malignant classifications in the dataset to minimize missed diagnoses.

Overall, the model performs better on Vindr-Mammo compared to the data distribution is more balanced CBIS-DDSM. We believe this may be due to Vindr-Mammo having approximately three times the data volume and being collected more recently, benefiting from advancements in imaging technology that enhance image quality. Therefore, addressing the impact of long-tail distribution and exploring the performance of these methods on larger and more balanced datasets will be an interesting direction for future research.

\textbf{Possibility of Weakly Supervised Lesion Localization Methods:} In this regard, we visualized the weakly supervised lesion localization performance of our model in Figure 5 and provided a more detailed comparison of different paradigms in the localization process in Figure 6. The results were analyzed and evaluated by clinicians, who acknowledged their effectiveness.

Figure \ref{fig_mom2} (a) presents a well-performing example, where most of the tumor lesion area is covered by the model-predicted patch-level images. This indicates that the image information of the lesion region has been effectively captured by the model and integrated through patch-level fusion into features that well reflect the classification outcome.
Interestingly, in Figure \ref{fig_mom2} (b), our model not only accurately localized the lesion but also showed activation in breast skin thickening in the fourth patch-level image. Additionally, in the second patch-level image of Figure \ref{fig_mom2} (c), the model detected a low-density small calcification that was not labeled in the data. This indicates that the model has not merely fit the data distribution but has effectively learned the representational information of breast images. However, despite this success, challenges remain. In Figure \ref{fig_mom2} (f), there is an issue with an insufficient number of patch-level images. Fortunately, Figure \ref{fig_mom2} (g), which is another view of the same breast from the same patient, shows that the predicted patch-level images cover almost all lesion areas, highlighting the advantage of multi-view information fusion. 
In addition to the issue of insufficient patch-level images, the patch-level image stacking problem in Figure \ref{fig_mom2} (b) also reflects the challenge. Designing a method for dynamically planning the number and size of patch-level images may be a promising direction for future exploration.
Additionally, a benign microcalcification was found in the fourth patch-level image of Figure \ref{fig_mom2} (n), which could potentially serve as a cue for clinicians.

With table \ref{Table5}, we observed that selecting the larger Res34-based model, compared to Res18-based, resulted in poorer performance. This may be due to the fact that larger models often require more complex optimization processes, potentially leading to unstable training or convergence to suboptimal solutions. The improved performance of the SwinT-based model also rules out the possibility of overfitting the training data due to excessive model parameters. This necessitates more effective training strategies to fully leverage the potential of large-parameter models.

\textbf{AI assisted medicine:}Exploring ways to better assist clinicians in screening tasks to reduce their workload also represents a more promising research avenue. AI technologies can significantly enhance the accuracy and efficiency of mammography screening, aiding radiologists in detecting early signs of breast cancer that might be missed by the human eye. By analyzing mammograms with sophisticated algorithms, AI can identify subtle patterns and anomalies, such as microcalcifications and masses, that indicate potential malignancies. These technologies have the potential to assist clinicians in identifying patterns, diagnosing conditions, and prioritizing cases more effectively, ultimately allowing them to focus on more complex aspects of patient care. Furthermore, AI can assist in standardizing mammogram interpretations, reducing variability between different radiologists and ensuring consistent diagnostic quality. This is particularly beneficial in low-resource settings, where training and expertise may vary widely. By integrating AI into the mammography screening process, healthcare systems can enhance their capacity to deliver timely and accurate breast cancer screening, ultimately reducing the burden of the disease. Therefore, investing in research that explores the integration of these tools in clinical settings is crucial for shaping the future of medicine.

\section{Conclusions}
The proposed Context Clustering Network with triple information fusion offers a promising solution to the challenges of breast cancer detection through mammography. By effectively integrating global, feature-based local, and patch-based local information, our approach addresses the limitations of traditional neural network architectures, such as the resolution of the mammography images to be processed is excessively high and the loss of microcalcifications and subtle structures due to down-sampling. The method’s computational efficiency and ability to associate structural and pathological features make it particularly suitable for clinical applications in mammography. Rigorous evaluation on public datasets Vindr-Mammo and CBIS-DDSM demonstrates the statistical robustness and superior performance of our approach, achieving significant improvements in AUC compared to existing methods. 

These results highlight the potential of our framework as a scalable and cost-effective tool for enhancing the accuracy and efficiency of large-scale mammography screening. By facilitating more precise and efficient breast cancer detection, our approach ultimately contributes to better patient outcomes and advances in healthcare delivery. Furthermore, the adaptability of our framework suggests its applicability to other high-resolution medical imaging tasks, paving the way for broader impacts in the field of medical diagnostics.

\section{Acknowledgments}
The author would like to acknowledge the funding support from the Clinical Research Program of the First Affiliated Hospital of Shenzhen University (2023YJLCYJ019) for this work.

\bibliographystyle{ieeetr}
\bibliography{ref}

\end{document}